\documentclass[usenatbib]{mn2e}
\usepackage{graphicx}

\title[GOES12 Remote Sounding]{Site testing at astronomical sites: seeing evaluation from satellite based data}

\author[S. Cavazzani et al.]{S. Cavazzani$^{1}$\thanks{E-mail:stefano.cavazzani@unipd.it}, S. Ortolani$^{1}$,
 V.Zitelli $^{2}$\\
$^{1}$Department of Astronomy, University of Padova, Vicolo
dell'Osservatorio 3, I-35122, Padova, Italy\\
$^{2}$INAF-Osservatorio Astronomico di Bologna, via Ranzani 1, I-40127, Bologna, Italy\\
}

\begin{document}

\date{Accepted 2011 October 04.  Received 2011 October 04; in original form 0000 Month 00.}

\pagerange{\pageref{firstpage}--\pageref{lastpage}} \pubyear{2009}

\maketitle

\label{firstpage}

\begin{abstract}

We present for the first time a new method to estimate the seeing using remote sounding from the IR  night time data of the GOES 12 satellite. We discuss the derived correlation between the ground data and the satellite derived values from the analysis of the sites located at Cerro Paranal (Chile) and Roque de los Muchachos (Canary Islands, Spain). We get a ground-satellite correlation percentage of about 90\%. Finally, studying the correlation between the afternoon data and the following night, we are able to provide a forecast of the photometric night quality.

\end{abstract}

\begin{keywords}
 atmospheric effects -- site testing -- methods: statistical.
\end{keywords}

\section{Introduction}

\begin{figure*}
  \centering
  \includegraphics[width=16cm]{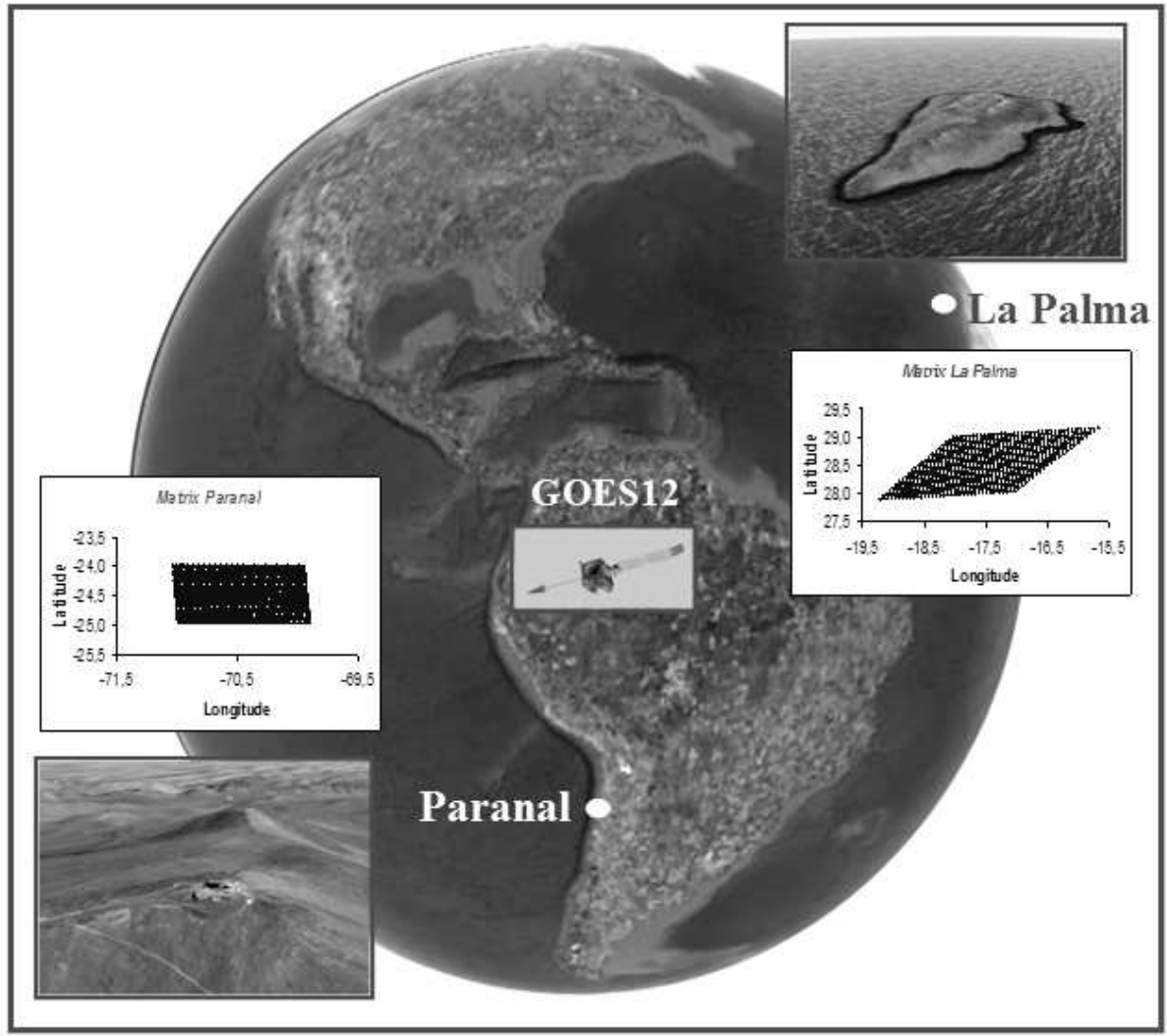}
  \caption{Location of the two sites involved in the analysis. As seen in the inserts the selected sites presents very different topographic conditions: La Palma is a sharp island, Paranal is isolated peaks over the Atacama's desert. The position of GOES12 satellite projected on the map. Figure also shows the comparison of one matrix at Paranal and La Palma. The deformation is a result of the satellite observation angle.}
             \label{terra}
\end{figure*}

The capability to optimize the scientific requirements to the observing conditions is a challenging effort crucial to improve the performances and to increase the final efficiency of the system telescope-instrumentation, mainly for very large telescopes. The first parameter needed for this goal is the knowledge of the usable nights. In the last decades this evaluation suffered of biases due to personal judgements, because they were based on visual inspection. A great improvement has been obtained with the use of satellite data. 
The second important parameter in the site selection and in the site characterization, is the image quality because, as well known, it affects the scientific quality of the results in many fields of the astronomical research.  Since the first campaigns for the site selection, the criteria were based simply on the direct analysis of the size of the stellar images. Now, with the progress of the knowledge in this area, we know that the seeing is characterized by multiple parameters and affected or simply linked to  several local and wide scale conditions, such as the external air temperature and gradients (Lombardi et al., (\cite{lombardi06}), hereafter Paper I), pressure, wind velocity (Lombardi et al., (\cite{lombardi07}), hereafter Paper II) and a link between these parameters and the optical turbulence (Cavazzani et al. (\cite{cava11})).
It is also crucial to know the evolution of the seeing with the time in short and long time scale, mainly for the future giant telescopes, for the optimization of the flexible scheduling.
In general the testing campaigns of the past were expensive and time consuming and limited to a few preselected sites. The use of the archive satellite data, instead, is of a great importance because it allows to simultaneously investigate several sites on a time base of many years.
A quantitative survey of cloud coverage and water vapor content  above several astronomical sites have been recently obtained using both satellite and ground based data by Erasmus and van Rooyen (\cite{erasmus06}), Erasmus and Sarazin (\cite{erasmus02}). They have been among the first to demonstrate the capability of the satellite data to give the amount of useful nights. Della Valle et al. (\cite{dellav10}, Paper III) used a similar analysis and, from independent data, found an agreement  of the amount of clear nights between satellite and ground based data at La Palma of about $80\%$. An evolution of this analysis is presented in Cavazzani et al. (\cite{cava11}) where we used a more sophisticated method and we introduced the concept of satellite stable night which is the best approximation of the concept of photometric nights. In this paper we present for the first time an estimation of the seeing obtained using the satellite remote sounding. We analyze the correlation between ground based seeing and the satellite based seeing. This analysis is applied to two very important astronomical international sites such as Cerro Paranal (Chile) and Roque de Los Muchachos (La Palma, Canary Islands, Spain) in order to validate the code in two very different climatic and topographic conditions. The location of the two sites is presented in Fig. \ref{terra}.
La Palma and Paranal are two sites in which the astronomical community built several facilities thanks to the good sky condition, moreover the community is strongly interested to maintain a high performance of the instrumentation. For this reason several authors focused the attention in the characterization of these two sites (Murdin \cite{murdin85}, Sarazin \cite{sara04},  Varela et al. \cite{varela}, ecc.).\\
The ESO staff was the pioneer of this topic and the long record of data collected at Paranal are useful tools to analyse the connection between astrophysical and physical environmental conditions. Differences with La Palma microclimate have been discussed in Paper I, Paper II, and Paper III. Paper I shows a complete analysis of the vertical temperature gradients and their correlation with the astronomical seeing, Paper II shows an analysis of the correlation between wind
and astronomical parameters as well as the overall long term weather conditions at La Palma. A statistical fraction of clear nights from satellite has been derived in Paper III using a basic approach to test the ability of the satellite to select clear nights.\\
The main conceptual difference between Erasmus \& van Rooyen (\cite{erasmus06}) analysis and Paper III is that they used the radio sounding vertical profile temperature as absolute reference to be compared with the brightness IR temperature measured by the satellite, while we used relative deviations from the bulk of data to detect the presence of clouds. In particular we selected two bands sensitive to the clouds and plotted one band versus the other.
The calibration of the plot gives the statistical fraction of usable nights.  The use of the two bands separately is efficient to sense thick clouds, but presents some limits in case of partial coverage or thin clouds. For this reason we have refined the analysis introducing a new band sensitive to the local phenomena and introducing a mathematical code  to correlate the three bands.  This analysis discriminates with success changes in air masses showing also a first  connection with seeing variations, as presented in  Cavazzani et al. (\cite{cava11}).
In this paper, to better analyse the correlation between satellite reflectivity and ground based image quality at La Palma and Paranal, we have used ground and satellite based data sampling the year 2009. We have used GOES satellite, to have homogeneous results with the previous papers and easy to compare and discuss. 
Table \ref{ST} shows the geographic position and view angle of the satellite for each site.
The paper is organized as follows: 

\begin{itemize}
	\item in Section \ref{GBD} we describe the used ground based data,
	\item in Section \ref{goes12} we describe the satellite based data, 
	\item in Section \ref{satacqui} we describe the satellite acquisition procedure,
	\item in Section \ref{rsbm} we describe the mathematical used model,
	\item in Section \ref{acf} we describe the atmospheric correlation function,
	\item in Section \ref{sub} we describe the approach to detect small clouds and local perturbations,
	\item in Section \ref{satclass} we describe the night classification from satellite,
	\item in Section \ref{seeing} we describe the satellite seeing,	
	\item in Section \ref{hda} we describe the temporal forecasting seeing 
	\item in Section \ref{conc} we report the discussion of the results.
\end{itemize}

\section{ Ground Based Data}
\label{GBD}

In this analysis we have compared satellite data with the image quality in term of FWHM obtained using the differential image motion monitoring (DIMM) at the two sites. Data at the Observatorio del Roque de los Muchachos (ORM) are derived from the Robotic Differential Image Motion Monitor (known as RoboDIMM \footnote{See http://catserver.ing.iac.es/robodimm/}) of Isaac Newton Telescope (INT). 
The INT RoboDIMM, like all classical DIMMs, relies on the method of differential image motion of telescope sub-apertures to calculate the seeing Fried parameter $r_{0}$. RoboDIMM forms four separated images of the same star, and measures image motion in two orthogonal directions from which it derives four simultaneous and independent estimates of the seeing.
The data interpretation makes use of the Sarazin and Roddier's DIMM algorithm described in Sarazin \& Roddier (\cite{dimm}), based on
the Kolmogorov theory of atmospheric turbulence in the free atmosphere.
At the present we do not have other seeing data to check possible local differences. But la Palma is the only site having several DIMM distributed along the top of the mountain. We are planning to follow this analysis using all the available DIMM data  to check possible local differences for a better characterization of the site and to correlate direct measurements such as $C^{2}_{n}(h)$ with satellite seeing. 
The seeing data at Paranal are obtained measuring the seeing of the DIMM at VLT observatory. The file contains also measurements of the flux of a reference star, in this way the flux of the star can trace the presence of clouds.
The ground based classification of the night quality instead has been done using the night observing log of each telescope.

\begin{table}
 \centering
 \begin{minipage}{80mm}
  \caption{Geographic characteristics of the analyzed sites and GOES12 satellite. The view angle is obtained through the formula $\theta=\sqrt{(\Delta LAT)^{2}+(\Delta LONG)^{2}}$.}
   \label{ST}
  \begin{tabular}{@{}lcccc@{}}
  \hline

  site      &LAT.&      LONG. & Altitude & View Angle \\
            &    &            &  Km      &            \\
 \hline
 Paranal    &   $-24^{\circ}37'$  &  $-70^{\circ}24'$  &  $2.630$   &  $25^{\circ}00'$  \\
 La Palma   &   $+28^{\circ}45'$  &  $-17^{\circ}52'$  &  $2.363$   &  $64^{\circ}10'$   \\
 \hline
 GOES12     &   $+0^{\circ}00' $  &  $-75^{\circ}00'$  & $35800$    &                   \\
 \hline

\end{tabular}
\end{minipage}
\end{table}

\section{Satellite Based Data}
\label{goes12}

In this analysis we have used GOES satellite because it is able to detects the infrared (IR) night time emitted radiation permitting to compare in a simultaneous way ground and satellite data. A detailed discussion of the performance of this satellite is presented in Cavazzani et al. (\cite{cava11}). The main advantage of GOES with respect other satellite is that GOES is able to observe the full Earth disk and it have on board an imager with five channels allowing the collection of five simultaneous images of almost half of the Earth hemisphere. Our choice to use the IR channels is because they allow the detection of the thermal radiation
emitted during the night from different atmospheric layers and/or from the soil. An appropriate choice of the wavelength allows to
choose the optimal layer emission height above the site. If it occurs well above the soil surface, the signal becomes
independent of the specific soil properties and of low level conditions. Phenomena occurring below the selected site (fog, low
clouds...) are also avoided. In some sites, for example at La Palma, this aspect is of crucial importance.

\section{Satellite data acquisition}
\label{satacqui}

For the purposes of this work we used GOES 12 equipped with the imager and we have analysed the year 2009. We have selected the  water vapour channel (B3 band) centred at $6.7~\mu m$, able to detect high altitude cirrus clouds, the infrared channel (B4 band) centred at $10.7~\mu m$, able to detect middle level clouds, and the $CO_2$ band (B6 band) centred at $13.3~\mu m$, able to sense small particle such as fog, ash and semi-transparent high clouds.
The selection of the IR channels was done in order to detect clouds at different heights, because water vapor absorbs electromagnetic radiation and then re-emits it in various wavelength bands, in particular in the infrared region at $6-7~\mu m$.
Each obtained data is a measurements of thermal radiation emitted during the  night by the Earth and received by the satellite detector. If clouds are not present, the emissions at  $10.7~\mu m$ reaching the satellite is largely not absorbed  by the atmosphere so the measured radiance values are due to emission from ground surface. Instead when clouds are present, the emissivity drops because is blocked.  Data are prepared by the Comprehensive Large Array-data Stewardship System (CLASS)\footnote{www.class.ngdc.noaa.gov} and processed using the free ware software McIDAS-V-1.0-beta4.
For each site we have identified and extracted  a sub-image of $1^\circ \times 1^\circ$ having the central
pixel close to the  coordinates given in Table \ref{ST}.
The use of the matrix is justified by the high correlation with the single pixel. In Cavazzani et al.(\cite{cava11}) we describe this correlation, in this paper we report the correlation coefficient values for 2009:

\begin{itemize}
	\item Matrix correlation coefficient at Paranal$=0.97$;
	\item Matrix correlation coefficient at La Palma$=0.93$;
\end{itemize}

The use of the matrix reduces the satellite noise and also allows us to observe the entire sky above the site.
Table \ref{BAND} shows the main characteristics of the selected bands.
For each night we have extracted the observations at different hours in local time: at 17:45, 20:45, 23:45, 02:45, 05:45, 7:45, 8:45 and 9:45. The evaluation of the amount of useful hours is done using all the night but 17:45 and 9:45. We have used both the brightness temperature at 17:45 and 9:45 to check a possible day-night correlation.   The last column of Table \ref{ST} shows the satellite view angle. The insert in Figure \ref{terra} shows the two different projections obtained from each acquisition at La Palma and Paranal.

\begin{table}
 \centering
 \begin{minipage}{80mm}
  \caption{Characteristics of the GOES12 used bands and resolution at Nadir.}
   \label{BAND}
  \begin{tabular}{@{}lcccc@{}}
  \hline
                & Window  & Passband          & Resolution      \\
                &         & $[\mu m]$         & [km]           \\
 \hline
 \textit{BAND3} & $H_{2}O$ & $6.50\div7.00$   & $4$\\
 \textit{BAND4} & $IR$     & $10.20\div11.20$ & $4$\\
 \textit{BAND6} & $CO_{2}$ & $13.30.$         & $8$\\
 \hline

\end{tabular}
\end{minipage}
\end{table}

\section{The code}
\label{rsbm}

In Cavazzani et al. (\cite{cava11}) there is an exhaustive description of the mathematics approach. We summarize here the main definition for a complete exposition.\\ The emitted monochromatic radiation intensity at a given $\lambda$ and
along a vertical path at the top of the atmosphere, incident at a satellite instrument is given by:

\begin{equation}
	R_{\lambda}=(I_{0})_{\lambda}\tau_{\lambda}(z_{0})+\int^{\infty}_{z_{0}}B_{\lambda}\left[T(z)\right]K_{\lambda}(z)dz
	\label{eq:a}
\end{equation}

where:

\begin{itemize}
	\item $K_{\lambda}(z)=\frac{d\tau_{\lambda}(z)}{dz}\Rightarrow$ Weighting Function (WF)
	\item $B_{\lambda}\left[T(z)\right]\Rightarrow$ Planck function profile as function of vertical temperature profile T
	\item $(I_{0})_{\lambda}\Rightarrow$ Emission from the earth surface at height $z_{0}$
	\item $\tau_{\lambda}(z)\Rightarrow$ Vertical transmittance from height $z$ to space
\end{itemize}

For a viewing path through the atmosphere at angle $\theta$ to the vertical, we have:

\begin{equation}
	\tau_{\lambda}(z,\theta)=e^{-sec\theta\int^{\infty}_{z}\kappa_{\lambda}(z)c(z)\rho(z)dz}	
\end{equation}

where:

\begin{itemize}
	\item $\rho(z)\Rightarrow$ Vertical Profiles of Atmospheric Density
	\item $\kappa_{\lambda}(z)\Rightarrow$ Absorption Coefficient
	\item $c(z)\Rightarrow$ Absorbing Gas Mixing Ratio
\end{itemize}

The emitted radiation intensities in each considered band  ${\lambda}$ are then:

\[	 R_{\lambda_{3}}=(I_{0})_{\lambda_{3}}\tau_{\lambda_{3}}(z_{0})+
\int^{\infty}_{z_{0}}B_{\lambda_{3}}\left[T(z)\right]K_{\lambda_{3}}(z)dz
\]
	
	\[	 R_{\lambda_{4}}=(I_{0})_{\lambda_{4}}\tau_{\lambda_{4}}(z_{0})+
\int^{\infty}_{z_{0}}B_{\lambda_{4}}\left[T(z)\right]K_{\lambda_{4}}(z)dz
\]

		\[	 R_{\lambda_{6}}=(I_{0})_{\lambda_{6}}\tau_{\lambda_{6}}(z_{0})+
\int^{\infty}_{z_{0}}B_{\lambda_{6}}\left[T(z)\right]K_{\lambda_{6}}(z)dz
\]

Each considered band is characterized by a weighting function (WF) giving the variation of the efficiency of the system as a function of the height. The peak of the efficiency specifies the layer from which the radiation is emitted and than the region of the atmosphere which can be sensed from space at fixed $\lambda$.\\ Assuming a standard atmosphere GOES12 WFs assign the following median height values to each band\footnote{See http://cimss.ssec.wisc.edu/}:

\begin{itemize}
	\item BAND3: $K_{\lambda_{3}}(z)=\frac{d\tau_{\lambda_{3}}(z)}{dz}\Rightarrow \approx 8000m$
	\item BAND4: $K_{\lambda_{4}}(z)=\frac{d\tau_{\lambda_{4}}(z)}{dz}\Rightarrow \approx 4000m$
	\item BAND6: $K_{\lambda_{6}}(z)=\frac{d\tau_{\lambda_{6}}(z)}{dz}\Rightarrow \approx 3000m$
\end{itemize}

The elevation assigned by the WF depend on the location of the selected sites but we can assume that these value can be assigned to both the interested sites having an heights ranging between 2 and 3 Km.

\section{Atmospheric Correlation Function}
\label{acf}

Instead to use each band separately in this analysis we have introduced a code to correlate the three bands. The correlation function $F_{C.A.}(t)$ is given by the equation (\ref{eq:b}):

\begin{equation}
F_{C.A.}=I_{\lambda_{3}}-[I_{\lambda_{6}}-I_{\lambda_{4}}] 
\label{eq:b}
\end{equation}

In mathematical terms this model provides the brightness temperature of the B3, B4 and B6 combination, given by equation:

	\[F_{C.A.}=\frac{R_{\lambda_{3}}+R_{\lambda_{4}}-R_{\lambda_{6}} }{\tau(z_{0})}+
\]

\begin{equation}
-\frac{\int^{\infty}_{z_{0}}B_{\lambda_{3}}[T(z)]K_{\lambda_{3}}+B_{\lambda_{4}}[T(z)]K_{\lambda_{4}}-B_{\lambda_{6}}[T(z)]K_{\lambda_{6}}dz}{\tau(z_{0})}	 
\label{eq:c}
\end{equation}

\begin{figure}
  \centering
  \includegraphics[width=8.5cm]{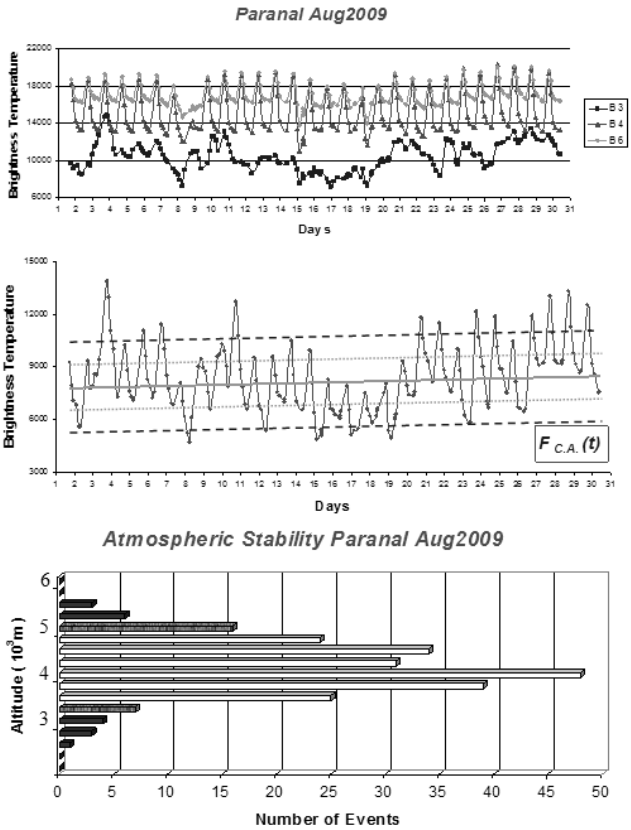}
  \caption{Atmosferic Correlation Function at Paranal, August 2009. Top figure shows the monthly plot of the three used bands.The central plot shows the $F_{C.A.}$ of August, the solid gray line is the $F_{C.A.}(t)$ linear regression. The brightness temperature is expressed in number of satellite counts as extracted with McIDAS-V program. The bottom part shows the distribution of the clear and stable nights as a function of the sensed height.}
             \label{pa}
   \end{figure}

The physical meaning of this model is the brightness temperature of the atmosphere reaching the satellite sensor as provided by the combination of the B3, B4 and B6 bands and it is  given by equation (\ref{eq:c}).
Therefore $F_{C.A.}$, provides information about the atmospheric evolution of the surveyed site. 
Moreover $F_{C.A.}$ provides information in both the height and quality of the perturbations over the surveyed site, that are both a function of the T brightness. In fact Figure \ref{pa} shows the obtained $F_{C.A.}(t)$ at Paranal for August 2009. 
Top Figure \ref{pa} shows the monthly trend of the three used bands. The central plot shows the $F_{C.A.}$ of August obtained from the three bands, the solid gray line is the $F_{C.A.}(t)$ linear regression. The bottom part shows the distribution of the clear and stable nights
discussed in Section \ref{satclass}.

\section{Detection of subtle phenomena}
\label{sub}

We describe here two different algorithm introduced to detect perturbations in two cases: low level perturbations located spatially very close to the telescope or very far to the telescope (located at the wedge of the matrix area) that means to have an incoming perturbation.

\subsection{Detection of small clouds in the matrix area}

\begin{figure}
  \centering
  \includegraphics[width=8.5cm]{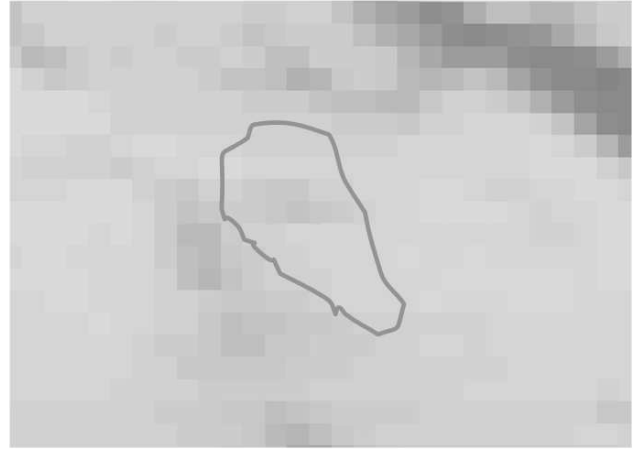}
  \caption{Example of a low standard deviation of pixel array.}
             \label{devia}
   \end{figure}

\begin{figure}
  \centering
  \includegraphics[width=8.5cm]{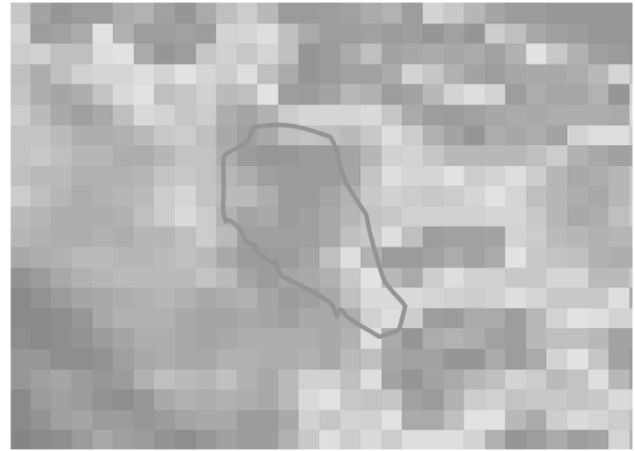}
  \caption{Example of a high standard deviation of pixel array.}
             \label{devia1}
   \end{figure}

In Della Valle et al. \cite{dellav10} the reflectivity flux has been obtained from the pixel of the matrix centred close to the coordinates of the interested site. To the aim to reduce the instrumental noise and to looking a wider field of view, we decided to replace the 1 pixel flux reflectivity with the mean value of the 1 degree matrix even centred at the coordinates of the interested site, moreover, to better discriminate small clouds distributed in the matrix area, that are missed giving a  limitation of the model as described in the previous section, we computed the standard deviation of each matrix.  
 In fact a high standard deviation signifies the presence of perturbations in the wall area. We are able also to see  incoming clouds approaching to the edge of the matrix area  
Figures \ref{devia} and \ref{devia1} show two examples in which the average value of the matrix in both the figures correspond to clear nights, but the standard deviation of Figure \ref{devia1} is high, showing a non real flat distribution of the satellite counts. This is the case of incoming perturbation to the telescope site. Considering the standard deviation, we obtain the following classification:

\begin{itemize}

	\item Standard deviation$(T_{B})\leq2\sigma\Longrightarrow$ Clear
	\item Standard deviation$(T_{B})>2\sigma\Longrightarrow$ Subtle Phenomena
	
\end{itemize}

\begin{figure}
  \centering
  \includegraphics[width=8.5cm]{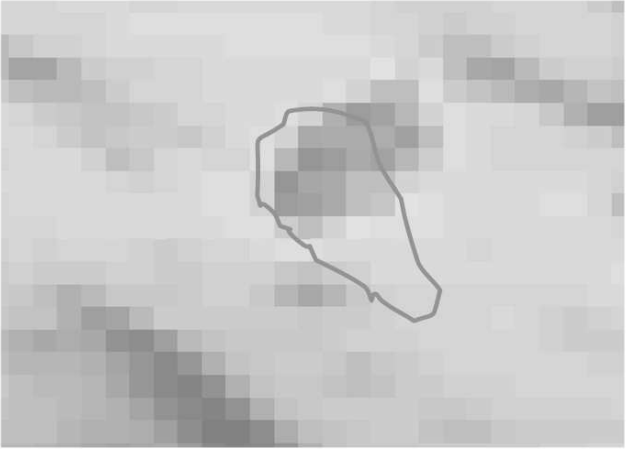}
\caption{Remote Sounding between the average and the single pixels: example of a large difference between the average and the single pixels.}
             \label{remote}
   \end{figure}

\subsection{Detection of local phenomena}

Finally, to better detect the presence of local phenomena close to the telescope, we introduce the difference between the mean matrix reflectivity and the single pixel reflectivity through the formula:

\begin{equation}
I_{RS}(Matrix/1Pixel)=\left|\overline{I_{\lambda_{4}}}-I_{\lambda_{4}(1Pixel)}\right|
\end{equation}

A high value of $I_{RS}$ shows the presence of a perturbation, even at low S/N level, not detectable using the simple standard deviation and the matrix average due the the high number of averaged pixels.\\
In particular, we use this mathematics classification: 

\begin{itemize}

	\item $\left|\overline{I_{\lambda_{4}}}-I_{\lambda_{4}(1Pixel)}\right|\leq2\sigma\Longrightarrow$ Clear
	\item $\left|\overline{I_{\lambda_{4}}}-I_{\lambda_{4}(1Pixel)}\right|>2\sigma\Longrightarrow$ Subtle Phenomena
	
\end{itemize}

\begin{figure}
  \centering
  \includegraphics[width=8.5cm]{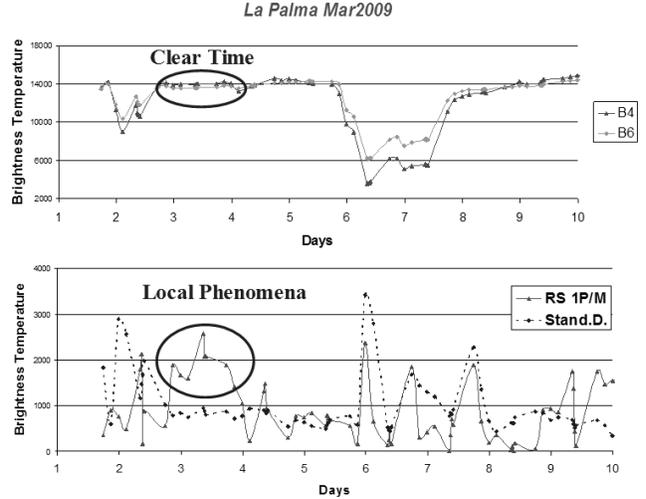}
  \caption{The B4 trend (upper panel) shows no  indicates while the $RS_{Matrix/1Pixel}$ (bottom panel)
   indicates the presence of local phenomena: the logbooks in fact describe the presence of high humidity and ice. The brightness temperature is expressed in number of satellite counts as extracted with McIDAS-V program.}
             \label{rs1}
   \end{figure}

\begin{figure}
  \centering
  \includegraphics[width=8.5cm]{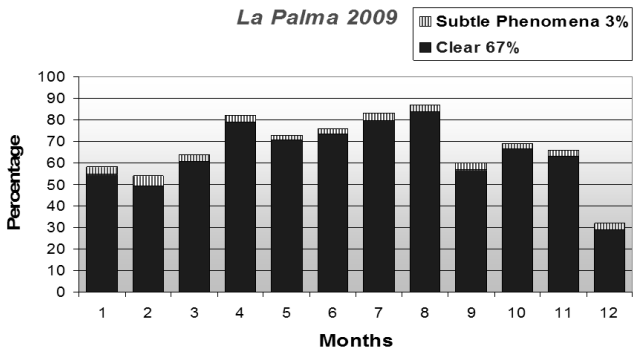}
  \caption{Subtle Phenomena. La Palma 2009.}
             \label{sub1}
   \end{figure}

Figure \ref{remote} represent one example of average corresponding to clear nights.
In this case, the $Matrix/1Pixel$ RS shows the presence of local stationary phenomena
not detected by the mean value of the matrix.
By the use of both the $Matrix/1Pixel$  and the standard deviation of each data we can better to detect local phenomena and thin clouds as shown in Figure \ref{rs1} in La Palma. In fact you see that the plot of B4 band evidenced by the circle is flat, typical of clear sky, the bottom of Figure \ref{rs1} plotting ${Matrix/1Pixel}$ difference show variations indicating the presence of local phenomena. A check with the logbooks describe the presence of high humidity and ice in this nights. We stress that these cases are rare,
in fact table \ref{SP}  shows the statistical result of this analysis for the 2009 at Paranal and La Palma. It is given the mean monthly percentage of clear nights and the fraction of the clear nights with low level phenomena. We see that only the 1\% of the 91\% of clear night at Paranal  is affected by low level phenomena, to compare with the 3\% of the 67\% at La Palma. In both cases is a very low number. At Paranal May shows an high number of $SUB_{P}$ phenomena. The check with the log gives high wind value coming form the see, that means high humidity justifying the high satellite value. Figure \ref{sub1} shows the amount of clear time at La Palma for the 2009, in gray it is shows the percentage of subtle phenomena.

\begin{figure}
  \centering
  \includegraphics[width=8.5cm]{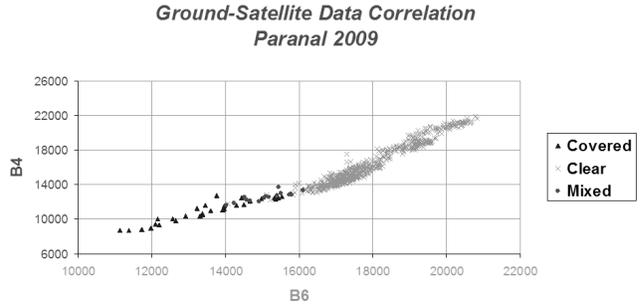}
  \caption{Temporal distribution of GOES12 B4 and B6 band emissivity  at Paranal in  2009.
  Sky quality classification has been carried out using the Paranal log.}
             \label{cor1}
   \end{figure}

\begin{figure}
  \centering
  \includegraphics[width=8.5cm]{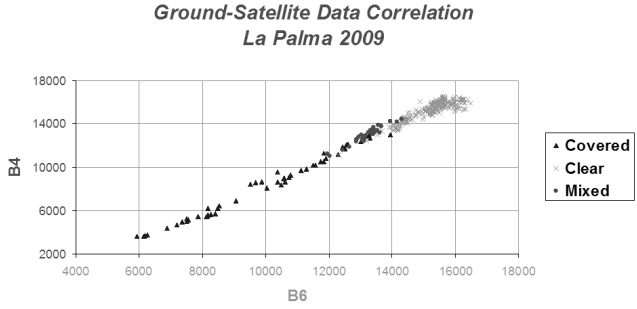}
  \caption{Temporal distribution of GOES12 B4 and B6 band emissivity  at La Palma in  2009.
  Sky quality classification has been carried out using the La Palma log.}
             \label{cor2}
   \end{figure}

\begin{table}
 \centering
 \begin{minipage}{80mm}
  \caption{Satellite Mean Monthly Percentage. Paranal and La Palma 2009. Subtle Phenomena ($SUB_{P}$).}
   \label{SP}
  \begin{tabular}{@{}llllccccccccc@{}}
  \hline
           & \multicolumn{2}{c}{Paranal}& \multicolumn{2}{c}{La Palma}\\
           & Clear Time& $SUB_{P}$ & Clear Time& $SUB_{P}$\\

 \hline
 January     &99.3 &1.0  &58.3 & 3.2 \\
 February  & 99.4 &1.5  &53.7 & 4.9 \\
 March     &97.0 & 1.3  &64.2 & 3.2 \\
 April     &  96.8 & 0.9 & 82.0 & 3.0\\
 May       &  89.0 & 2.1 & 73.3 & 2.1\\
 June      &  80.1 & 0.9 & 75.9 & 2.5 \\
 July      & 79.5 & 1.6 & 82.9 & 3.3 \\
 August    &  90.5 & 1.8 & 86.5 & 3.0\\
 September & 85.3 & 0.3 & 60.2 & 2.2\\
 October   &  85.7 & 0.7 & 69.2 & 2.5 \\
 November  & 99.2 & 0.0 & 66.2 & 2.8 \\
 December  &  98.3 & 0.4 & 32.2 & 3.0 \\
\hline
 Mean      & 90.8 & 1.0 & 67.1 & 3.1 \\
 \hline

\end{tabular}
\end{minipage}
\end{table}

\section{Temporal Satellite classification}
\label{satclass}

To have reliable prediction of the time quality, we have used a high temporal resolution using for each night the following series of data: 20:45, 23:45, 2:45, 5:45, 7:45, 8:45.
Using the brightness temperature obtained for each considered hours we have obtained the monthly atmospheric correlation function. 
Figures \ref{cor1} and \ref{cor2} show the plot of the obtained temporal emissivity of B4 band vs B6 for the 2009 at Paranal and La Palma. Nights are classified according to the comment of the observing logs. 
Clear time presents high values of emissivity at both sites. As in Paper III, the classification of satellite time quality is done assuming that the maximum monthly brightness temperature in B4 band ($T^{Max}_{B}$) occurs in clear condition. The other brightness hourly temperatures are correlated with  $T^{Max}_{B}$ when:

\begin{itemize}

	\item $T^{Max}_{B}-T_{B}\leq2\sigma\Longrightarrow$ Clear
	\item $2\sigma<T^{Max}_{B}-T_{B}\leq3\sigma\Longrightarrow$ Mixed
	\item $T^{Max}_{B}-T_{B}>3\sigma\Longrightarrow$ Covered
	
\end{itemize}

where $T_{B}\Rightarrow$ Brightness temperature of the $1^\circ\times1^\circ$ matrix.
Table \ref{delta} shows the obtained percentage of clear, mixed and covered nights at Paranal and at La Palma for the year 2009 using all the algorithm previous described. The ground based classification is derived from the comments of the night logbook. We found a very good agreement in both the two sites between ground and satellite data.
The last row of Table \ref{delta} shows the percentage of accuracy to associate to each obtained fraction of nights. The uncertainty is computed as follows:

\begin{itemize}
	\item $\Delta_{Clear/Mixed}\Rightarrow$ Clear/Mixed Uncertainty
	\item $\Delta_{Clear/Covered}\Rightarrow$ Clear/Covered Uncertainty
	\item $\Delta_{Mixed/Covered}\Rightarrow$ Mixed/Covered Uncertainty
\end{itemize}

Usually the quality of the ground based clear nights is divided between photometric and spectroscopic nights. Also for the satellite classification we have introduced a similar definition introducing the concepts of stable night (photometric) and clear night (spectroscopic).
Considering the value of the $F_{C.A.}(t)$ linear regression $T^{Trendline}_{B}$ we define:

\begin{itemize}
	\item $\left|T_{B}-T^{Trendline}_{B}\right|\leq\left|1\sigma\right|\Longrightarrow$ Stable
	\item $\left|1\sigma\right|<\left|T_{B}-T^{Trendline}_{B}\right|\leq\left|2\sigma\right|\Longrightarrow$ Clear
	\item $\left|T_{B}-T^{Trendline}_{B}\right|>\left|2\sigma\right|\Longrightarrow$ Covered
	
\end{itemize}

where:

\begin{enumerate}
  \item $T^{Trendline}_{B}\Rightarrow$ Brightness temperature of the $F_{C.A.}(t)$ linear regression computed in one month
	\item $T_{B}\Rightarrow$ Brightness temperature of the $1^\circ\times1^\circ$ matrix in one hour
\end{enumerate}

Table \ref{Meanpa}  shows the obtained satellite mean monthly percentage of clear and stable time at Paranal and la Palma.

\begin{table*}
 \centering
 \begin{minipage}{170mm}
  \caption{Temporal data analysis of Clear/Mixed/Covered time at Paranal and La Palma in 2009.}

   \label{delta}
  \begin{tabular}{@{}llllccccccc@{}}
  \hline
  &   \multicolumn{3}{c}{Ground} & \multicolumn{3}{c}{Satellite}\\
     & Clear& Mixed & Covered& Clear& Mixed& Covered\\

 \hline
 Paranal   & 90.1\% & 2.2\%    & 7.8\% & 90.8\% & 2.6\% & 6.6\% \\
 La Palma  & 65.8\%  & 5.0\%   & 29.3\% & 67.0\%   & 4.5\%   &   28.5\% \\
 \hline
  &  \multicolumn{3}{c}{Paranal} & \multicolumn{3}{c}{La Palma}\\
Uncertainty & $\Delta_{Clear/Mixed}$   &  $\Delta_{Clear/Covered}$   & $\Delta_{Mixed/Covered}$ & $\Delta_{Clear/Mixed}$   &  $\Delta_{Clear/Covered}$   & $\Delta_{Mixed/Covered}$  \\
\hline
Percentage &  $1.2\%$ &  $0.4\%$  &  $0.8\%$ &   $1.3\%$ &  $0.5\%$  &  $0.8\%$   \\
 \hline

\end{tabular}
\end{minipage}
\end{table*}

\begin{figure}
  \centering
   \includegraphics[width=8.5cm]{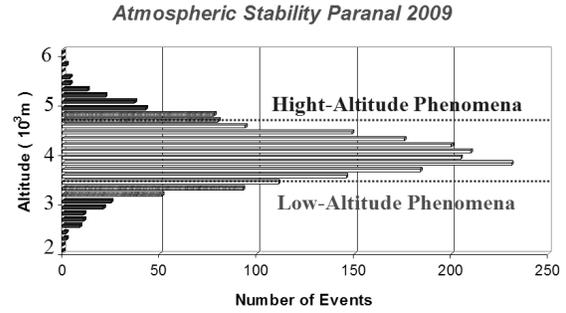}
  \caption{Histogram of annual atmospheric stability at Paranal. Light-gray bars represent
the stable nights, gray bars clear but unstable nights, black
bars the nights covered.}
             \label{s}
   \end{figure}

\begin{figure}
  \centering
   \includegraphics[width=8.5cm]{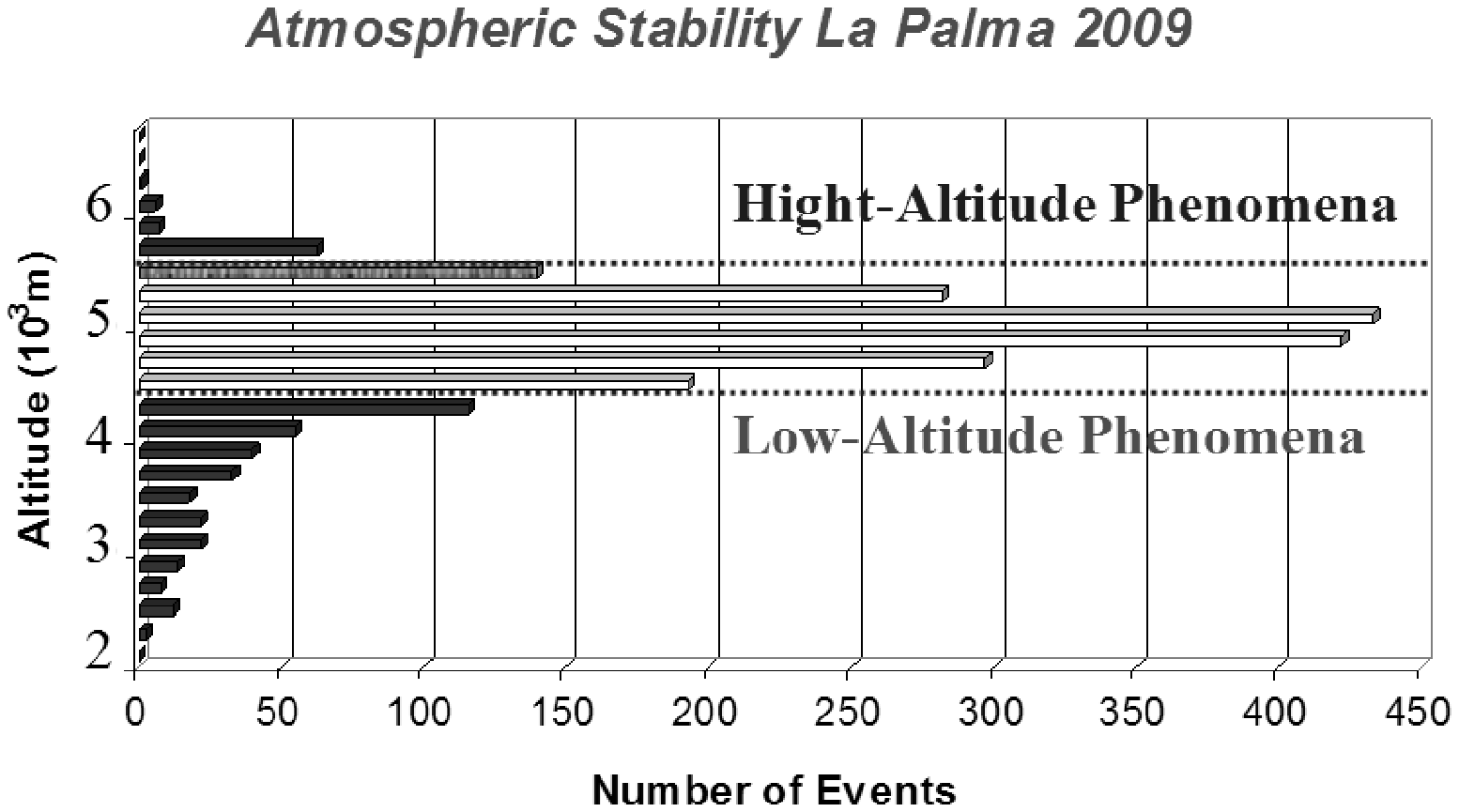}
  \caption{Histogram of annual atmospheric stability at La Palma. Light-gray bars represent
the stable nights, gray bars clear but unstable nights, black
bars the nights covered.}
             \label{s1}
   \end{figure}

\begin{figure}
  \centering
  \includegraphics[width=8.0cm]{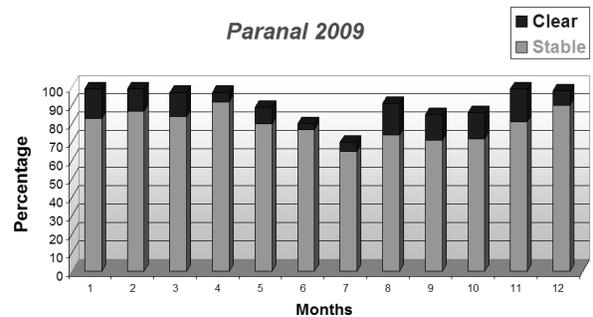}
  \caption{Clear and stable night fractions at Paranal 2009 from GOES12 satellite.}
             \label{pa1}
\end{figure}

Figures \ref{s} and \ref{s1} show the distribution of the amount of clear, stable and covered time at Paranal and la Palma for the year 2009 obtained from the $F_{C.A.}(t)$. The maximum of the distribution shows the sensed height and it gives the height in which occur the atmospheric phenomena. Figure \ref{pa1} shows the monthly distribution of the clear and stable nights at Paranal for the considered year. We see that during the winter months is low the percentage of stable time and the night is mostly only clear.

\begin{table}
 \centering
 \begin{minipage}{170mm}
  \caption{Satellite Mean Monthly Percentage for the year 2009.}
   \label{Meanpa}
  \begin{tabular}{@{}lllllccccccccccc@{}}
  \hline
           &\multicolumn{2}{c}{Paranal} & \multicolumn{2}{c}{La Palma}\\
 Paranal         & Clear Time& Stable & Clear Time& Stable\\

 \hline
 January   & 99.3 & 83.3 & 58.3 & 57.6\\
 February  & 99.4 & 86.6 & 53.7 & 48.8\\
 March     & 97.0 & 84.4 & 64.2 & 57.6\\
 April     & 96.8 & 92.2 & 82.0 & 78.8\\
 May       & 89.0 & 79.7 & 73.3 & 64.0\\
 June      & 80.1 & 76.7 & 75.9 & 72.3\\
 July      & 69.5 & 64.9 & 82.9 & 75.6\\
 August    & 90.5 & 74.0 & 86.5 & 76.0\\
 September & 85.3 & 70.9 & 60.2 & 56.8\\
 October   & 85.7 & 71.7 & 69.2 & 64.7\\
 November  & 99.2 & 80.8 & 66.2 & 59.2\\
 December  & 98.3 & 89.8 & 32.2 & 30.7\\
\hline
 Mean      & 90.8 & 79.6 & 67.1 & 61.6\\
\hline

\end{tabular}
\end{minipage}
\end{table}

\begin{table}
 \centering
 \begin{minipage}{80mm}
  \caption{Mathematical and statistical uncertainties of the model in 2009 at Paranal and La Palma.}
   \label{delta3}
  \begin{tabular}{@{}lccc@{}}
  \hline

 Site                 & $\Delta_{Total}$  & $N(G;S)$ & $\Delta_{Statistical}$  \\

 \hline
 Paranal              & $1.4\%$           & 1510      &  $0.05\%$               \\
 La Palma            & $1.5\%$           & 1510     &   $0.06\%$               \\
 \hline

\end{tabular}
\end{minipage}
\end{table}

\begin{table}
 \centering
 \begin{minipage}{80mm}
  \caption{Satellite FWHM at Paranal for the year 2009.}
   \label{fwhmsat}
  \begin{tabular}{@{}lllllccccccccccc@{}}
  \hline
                
 Months   & $FWHM^{Mean }_{Sat}$ & $FWHM^{Mean }_{Ground}$ & $COR_{Coef}$ \\

 \hline
 January   & 0.9 & 0.9 & 0.91 \\
 February  & 0.8 & 0.8 & 0.97\\
 March     & 0.8 & 0.8 & 0.88\\
 April     & 0.7 & 0.7 & 0.93 \\
 May       & 0.8 & 0.8 & 0.93 \\
 June      & 0.8 & 0.8 & 0.84 \\
 July      & 0.8 & 0.8 & 0.79 \\
 August    & 0.9 & 0.9 & 0.92\\
 September & 0.9 & 0.9 & 0.96\\
 October   & 0.8 & 0.9 & 0.84 \\
 November  & 0.8 & 0.9 & 0.95\\
 December  & 0.8 & 0.8 & 0.95 \\
\hline

\end{tabular}
\end{minipage}
\end{table}

\begin{table}
 \centering
 \begin{minipage}{80mm}
  \caption{Satellite FWHM at La Palma for the year 2009. We have calculated the correlation coefficients only for the months in which the RoboDIMM gives us values for more than ten nights. Moreover the mean seeing only refers to the clear time due to the fact that in the covered time the RoboDIMM does not work.}
   \label{fwhmsat1}
  \begin{tabular}{@{}lllllccccccccccc@{}}
  \hline
                
 Months   & $FWHM^{Mean }_{Sat}$ & $FWHM^{Mean }_{Ground}$ & $COR_{Coef}$ \\

 \hline
 January   & -   & -   & -   \\
 February  & -   & -   & -    \\
 March     & 0.9 & 1.0 & 0.89 \\
 April     & 1.0 & 1.0 & 0.91 \\
 May       & 0.8 & 0.8 & 0.94 \\
 June      & 0.9 & 0.8 & 0.93 \\
 July      & 0.9 & 0.9 & 0.92 \\
 August    & 0.8 & 0.7 & 0.92 \\
 September & 0.8 & 0.7 & 0.95 \\
 October   & -   & -   & -   \\
 November  & -   & -   & -   \\
 December  & -   & -   & -   \\
\hline

\end{tabular}
\end{minipage}
\end{table}

\section{Satellite Calculation of Seeing} 
\label{seeing}

In  Cavazzani et al.(\cite{cava11}) we shown that the adopted code is able to discriminate with success variations of the atmospheric stability function ($F_{C.A.}(t)$ ) with the optical turbulence showing the first  connection between $F_{C.A.}(t)$  and seeing. 
In this paper we are going deeper in this analysis and, to better analyse the correlation between satellite reflectivity and ground based image quality at La Palma and Paranal, we have used ground and satellite based data sampling the year 2009.  
In particular we introduce for the first time the concept of satellite seeing. 
The $F_{C.A.}(t)$  measure the temperature in different atmospheric layers, and as well as the ground based  $C^{2}_{T}$ is linked to the $r_{0}$ and to the FWHM, it is possible to derive a satellite based $C^{2}_{T}$, and consequently $C^{2}_{n}$. The zero point is given empirically in this analysis.
Using the basic formulae of the seeing theory such as Fried's radius $r_{0}$ we have:

\begin{equation}
\label{friede}
	r_{0}=\left[0.423\cdot\frac{4\pi^{2}}{\lambda^{2}}\cdot\frac{1}{cos(\theta_{zen})}\int C^{2}_{n}\cdot dz\right]^{-\frac{3}{5}}
\end{equation}

where $C^{2}_{n}$ is the refractive index structure parameter:

	\[C^{2}_{n}=\left[80\cdot10^{-6}\frac{P}{T}\right]\cdot C^{2}_{T}
\]

The full width at half maximum is given by the following formula:

\begin{equation}
	FWHM=0.98\frac{\lambda}{r_{0}}
\end{equation}

While the satellite FWHM is obtained through an our empirical model. 
If we assume:

	\[\left|T_{B}-T^{Trendline}_{B}\right|\propto C^{2}_{T}\propto C^{2}_{n}
\]

we can replace the $C^{2}_{n}$ value in the Equation \ref{friede} obtaining a satellite $r_{0}$ calculation:

\begin{equation}
\label{seeing3}
r_{0, Sat}=\left[0.423\cdot\frac{4\pi^{2}}{\lambda^{2}}\cdot\Lambda(\theta)\cdot\frac{\left|T_{B}-T^{Trendline}_{B}\right|}{z}\right]^{-\frac{3}{5}}	
\end{equation}

\begin{figure}
  \centering
  \includegraphics[width=8.0cm]{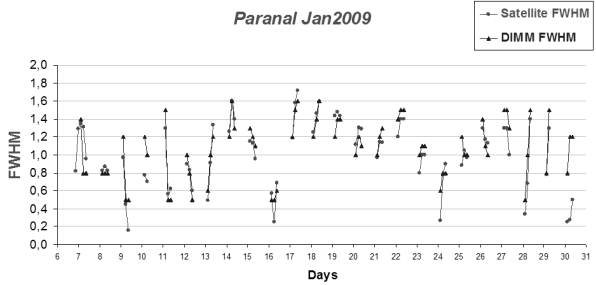}
  \caption{Ground Data-Satellite Data Correlation. Comparison between the FWHM calculated from the ground and the satellite FWHM. Paranal, January 2009 (Correlation Coefficient$=0.91$). The satellite FWHM is calculated through the Formula \ref{seeing1}.}
             \label{fwhms}
   \end{figure}

Finally, using this value we get the formula for satellite FWHM:

\begin{equation}
\label{seeing1}
FWHM_{Sat}=0.58\cdot\lambda^{-\frac{1}{5}}\cdot\left[4\pi^{2}\cdot\Lambda(\theta)\cdot\frac{\left|T_{B}-T^{Trendline}_{B}\right|}{z}\right]^{\frac{3}{5}}
\end{equation}

where $\Lambda(\theta)$ is an empirical constant defined by the formula:

\begin{equation}
	\Lambda(\theta)=\frac{10^{-12}}{cos\theta}
\end{equation}

where $\theta$ is the satellite angle of view.\\
Figure \ref{fwhms} shows the comparison between ground based FWHM and satellite based FWHM computed in the same hours.
We note the very good agreement between the two set of data.
Figure \ref{cc} shows the dispersion of this correlation and its linear regression.
A tentative physical interpretation of our correlation could be related to the Richardson number $R_{i}$ dependent on the vertical temperature gradient.
Tables \ref{fwhmsat} and \ref{fwhmsat1} show the comparison between the seeing as given by the ground and those computed by satellite using the equation \ref{seeing1}.
We make the following observations to discuss the obtained values: $FWHM^{Mean }_{Ground}$ values at Paranal are the DIMM data and not the VLT values;
at La Palma instead we have calculated the correlation coefficients only for the months in which the RoboDIMM gives us values for more than ten nights. Moreover the mean seeing only refers to the clear time due to the fact that the RoboDIMM does not work during cloudy nights.

\begin{table*}
 \centering
 \begin{minipage}{150mm}
  \caption{Forecast at Paranal and at La Palma for the year 2009. $A\rightarrow N$ is the correlation between the afternoon and the next night and $N\rightarrow M$ is the correlation between the morning and the night before.}
   \label{forecast}
  \begin{tabular}{@{}lllllccccccccccc@{}}
  \hline
   & &\multicolumn{2}{c}{Paranal} & \multicolumn{2}{c}{La Palma}\\              
 Months         & Days & $A\rightarrow N$ Correlation & $N\rightarrow M$ Correlation& $A\rightarrow N$ Correlation & $N\rightarrow M$ Correlation\\

 \hline
 January   &31& 100.0 & 100.0  & 93.5 & 96.8  \\
 February  &28& 96.4 & 100.0  & 85.7 & 92.9 \\
 March     &31& 96.8 & 100.0 & 90.3 & 93.5  \\
 April     &30& 96.7 & 100.0 & 90.0 & 96.7 \\
 May       &31& 100.0 & 96.8  & 100.0 & 96.8  \\
 June      &30& 100.0 & 93.3  & 90.0 & 90.0 \\
 July      &31& 93.5 & 93.5 & 96.8 & 96.8 \\
 August    &31& 96.8 & 96.8  & 96.8 & 100.0 \\
 September &30& 100.0 & 100.0 & 96.7 & 100.0 \\
 October   &31& 100.0 & 100.0 & 96.8 & 96.8 \\
 November  &30& 100.0 & 100.0  & 100.0 & 100.0 \\
 December  &31& 96.8 & 96.8 & 96.8 & 96.8\\
\hline

\end{tabular}
\end{minipage}
\end{table*}

\section{Temporal Forecasting Seeing  Analysis}
\label{hda}

In this section we analysed for the first time the possibility to give a forecasting value of the seeing a few hours before starting the observations. We have proceeded in two different ways to check the capability and the best procedure. 
In the first test we have correlated the brightness temperature obtained from the  value at 9:45 with the brightness temperature obtained using the values of the nights before.

\begin{figure}
  \centering
  \includegraphics[width=8.0cm]{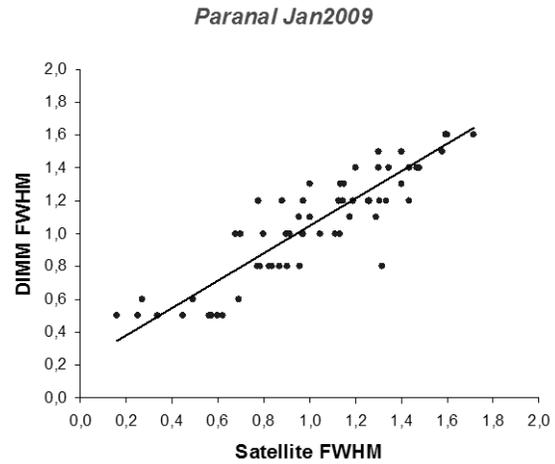}
  \caption{Ground Data-Satellite Data Correlation. Figure shows the dispersion of this correlation and its linear regression. Paranal, January 2009 (Correlation Coefficient$=0.91$). }
             \label{cc}
   \end{figure}

\begin{figure}
  \centering
  \includegraphics[width=8.0cm]{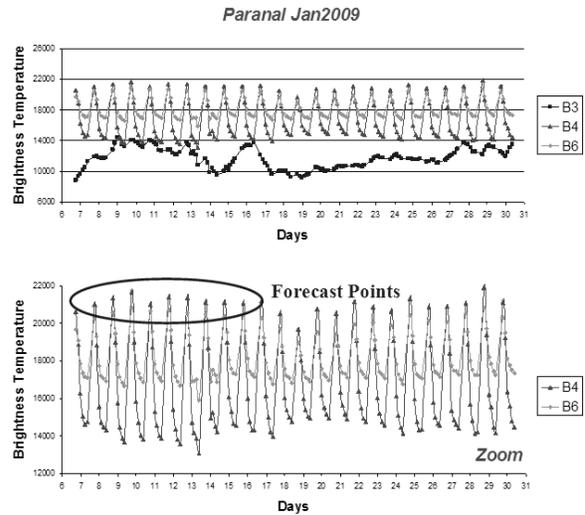}
  \caption{GOES 12 emissivity in B3, B4, B6 bands (upper panel) at Paranal for January 2009. Botton panel shows
  the B4, B6 vertical scale zoom. The brightness temperature is expressed in number of satellite counts as extracted with McIDAS-V program.}
             \label{sat1}
\end{figure}

\begin{figure*}
  \centering
  \includegraphics[width=16cm]{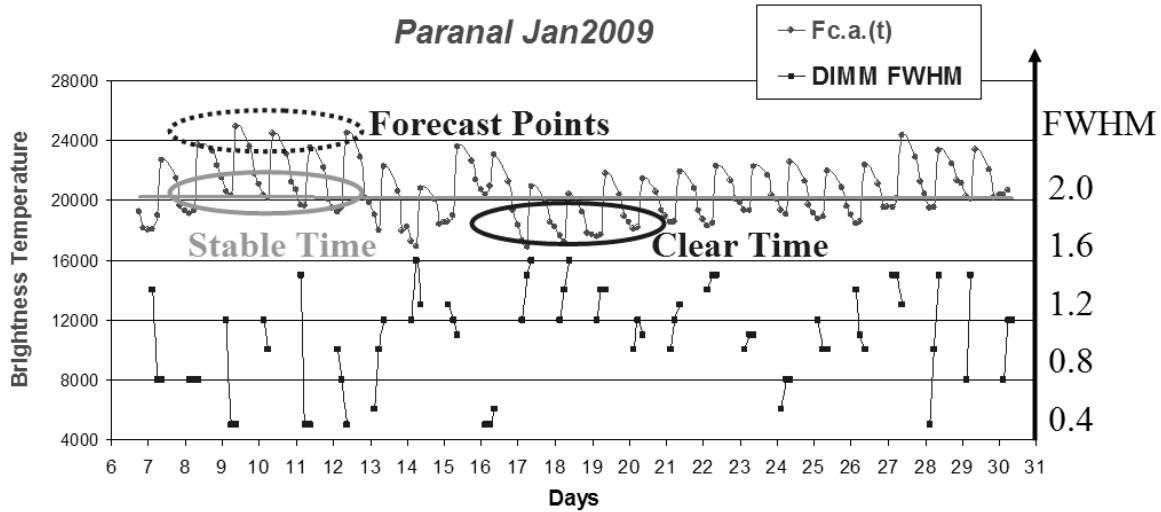}
  \caption{Figure shows the trend of the $F_{C.A.}(t)$. In this function we have highlighted the new points used for the night quality forecast (Forecast points). Through the position of these points we can predict whether the night will be stable (Stable time) or clear (Clear time). In fact, Figure also shows the DIMM FWHM values at Paranal (drawn in black). We note that at these stable time points we have low seeing values, conversely we have high seeing values at clear time points (The ordinate on the right shows the DIMM FWHM values). The brightness temperature is expressed in number of satellite counts as extracted with McIDAS-V program.}
             \label{fore}
   \end{figure*}

In the second test we have correlated the brightness temperature obtained from the afternoon value at 17:45 and the same night time. 
Figure \ref{sat1} shows the monthly distribution of the three bands at Paranal for January 2009 (upper panel) and the zoom (bottom panel) in which is possible to see the new day point used for the forecasting seeing. We see that in case of hight day values the  night after is stable (photometric night). It is interesting to note that in this analysis we are able to give a percentage of useful nights instead of useful time. 
Table \ref{forecast} shows the monthly values of the derived correlation at the two sites. Column 1 of tables shows the month, column 2 shows the number of used days, column 3 shows the afternoon to night correlation ($A\rightarrow N$ is the correlation between the afternoon and the next night) and  column 4 shows the night to morning correlation ($N\rightarrow M$ is the correlation between the morning and the night before). For our analysis we are interested to column 3 that give the correlation of all the available day-night data. We see that at Paranal and at La Palma the correlation decrease during the winter months.
Tables \ref{change} and \ref{change1} instead show the period of the day in which the meteorological variation occurred. Column 1 shows the month, column 2 shows the percentage of the variation occurred in the time range between 5 p.m. and 6 a.m., column 3 shows the variation occurred in the time between 9 p.m. and 10 a.m., column 4 show the percentage obtained for the not analysed day ($10_{a.m.}-5_{p.m.}$) and obtained for difference.\\
These numbers are obtained through the percentage of clear time (Table \ref{Meanpa}) and the correlation percentages (e.g. If we have clear time $=70\%\Longrightarrow$ covered time $=30\%$. Then we have $A\rightarrow N$ correlation $=95\%$ and $N\rightarrow M$ correlation $=90\%$. This means that $5\%$ of the meteorological changes occurred between 5 p.m. and 6 a.m., $10\%$ between 9 p.m. and 10 a.m. and the remaining $15\%$ between 10 a.m. and 5 p.m.). We see that most of the changing occur during the day time (from 10 a.m. to 5 p.m.) so it is possible to correlate the afternoon satellite data with the next night satellite data.

\begin{figure}
  \centering
  \includegraphics[width=8.0cm]{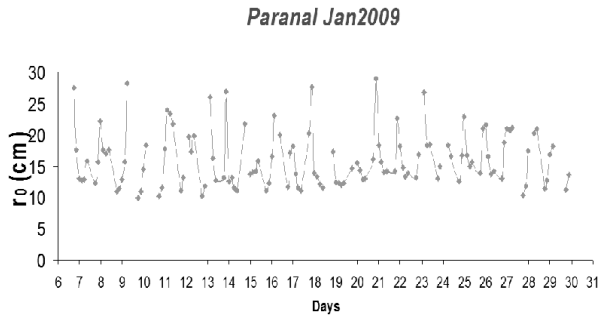}
  \caption{Fried's radius values calculated from satellite ($r_{0,Sat}$) at Paranal, January 2009. The satellite FWHM calculated with these values has a correlation coefficient of 0.91 with the ground FWHM (see Figure \ref{fwhms}).}
             \label{sat2}
\end{figure}

Figure \ref{fore} shows the trend of the $F_{C.A.}(t)$ obtained using all the brightness values and the DIMM seeing. The gray line is the best fit of the monthly plot. In this function we have highlighted the new points used for the night quality forecast (Forecast points). Through the position of these points we can predict whether the night will be stable (Stable time) or clear (Clear time).\\
Figure \ref{sat2} shows the obtained $r_{0}$ values from satellite.
These values are obtained through the model described in Section \ref{seeing} (Equation \ref{seeing3}).
Moreover the value of $r_{0}$ is computed taking into account the Paranal height and it is given in the visible range. We obtained values close to those obtained using ground based data.
In fact, the FWHM calculated by the $r_{0,Sat}$ value has a high correlation coefficient with the ground FWHM (see Tables \ref{fwhmsat}, \ref{fwhmsat1} and Figure \ref{cc}).\\

\section{Conclusion}
\label{conc}

In this paper, as first, we have introduced the concept of satellite seeing using remote sounding from the IR  night time data of the GOES 12 satellite. We have discussed the derived correlation between the ground data and the satellite derived values from the analysis of the sites located at Cerro Paranal (Chile) and Roque de Los Muchachos  (Canary Islands, Spain) for the 2009. In this analysis we used the $F_{C.A.}(t)$ obtained correlating the monthly mean values obtained in a 1 deg matrix of each of the three selected bands. This functions that is a measure of the gradient of temperature among the three layers sampled by the three bands is, as well as the ground based $C^{2}_{T}$, linked to the $r_{0}$. The $r_{0}$ values derived using $F_{C.A.}(t)$ (see Section \ref{seeing}) at Paranal and La Palma are close to ground based values, in particular the FWHM calculated by the $r_{0,Sat}$ value has a high correlation coefficient with the ground FWHM (see Tables \ref{fwhmsat}, \ref{fwhmsat1} and Figure \ref{cc}). In this first analysis we obtained empirically the zero point using the DIMM seeing from each site. We have demonstrate that the plot of the seeing from satellite is in good agreement with the DIMM seeing of the same month (see Figure \ref{fwhms}) showing a correlation ranging between 80\% and 97 \% during the months at Paranal. Figure \ref{cc} shows an example of the dispersion of this correlation and its linear regression for January 2009 at Paranal.
We found a better correlation at la Palma (89\% to 95\%), this is due to the fact that the correlation only refers to the clear time, in fact the RoboDIMM does not work during cloudy nights.

\begin{table}
 \centering
 \begin{minipage}{80mm}
  \caption{Meteorological changes at Paranal for the year 2009.}
   \label{change}
  \begin{tabular}{@{}lllllccccccccccc@{}}
  \hline
                
 Months        & $5_{p.m.}-6_{a.m.}$  & $9_{p.m.}-10_{a.m.}$ & $10_{a.m.}-5_{p.m.}$\\

 \hline
 January   & 0.0 & 0.0 & 100.0 \\
 February  & 100.0 & 0.0 & 0.0 \\
 March     & 100.0 & 0.0 & 0.0 \\
 April     & 100.0 & 0.0 & 0.0 \\
 May       & 0.0 & 29.3 & 70.7 \\
 June      & 0.0 & 33.3 & 66.7 \\
 July      & 21.5 & 21.5 & 57.0 \\
 August    & 35.8 & 35.8 & 28.3 \\
 September & 0.0 & 0.0 & 100.0 \\
 October   & 0.0 & 0.0 & 100.0 \\
 November  & 0.0 & 0.0 & 100.0 \\
 December  & 50.0 & 50.0 & 0.0 \\
\hline

\end{tabular}
\end{minipage}
\end{table}

\begin{table}
 \centering
 \begin{minipage}{80mm}
  \caption{Meteorological changes at La Palma for the year 2009.}
   \label{change1}
  \begin{tabular}{@{}lllllccccccccccc@{}}
  \hline
                
 Months        & $5_{p.m.}-6_{a.m.}$  & $9_{p.m.}-10_{a.m.}$ & $10_{a.m.}-5_{p.m.}$\\

 \hline
 January   & 15.4 & 7.7 & 77.0 \\
 February  & 31.1 & 15.5 & 53.4 \\
 March     & 26.9 & 17.9 & 55.2\\
 April     & 55.6 & 18.5& 25.9\\
 May       & 0.0  & 11.9 & 88.1 \\
 June      & 41.7 & 41.7 & 16.7 \\
 July      & 19.0 & 19.0 & 62.0 \\
 August    & 24.8 & 0.0 & 75.2\\
 September & 8.3& 0.0 & 91.7\\
 October   & 10.4 & 10.4 & 79.2 \\
 November  & 0.0 & 0.0 & 100.0 \\
 December  & 4.7 & 4.7 & 90.5 \\
\hline

\end{tabular}
\end{minipage}
\end{table}

Any comments we can gives about the obtained values due the zero point, but we have intention to refine the procedure.  As further step we are giving for the first time the forecasting seeing from satellite (see Figure \ref{fore}). We have proceeded in two ways to select the best procedure. In the first test we correlated the brightness temperature of the morning 9:45 with the values of the night before. In the second test we correlated the brightness temperature of the 17:45 afternoon with the night after. The two procedures seems to be show similar results, with a marginal higher percentage in the night-morning values for both the sites, but for the purpose of the prediction of the image quality for the incoming observing night we can use the correlation afternoon-night. We see that at Paranal the correlation decrease during the winter months, instead we found a more homogeneous distribution at la Palma.\\
Through this afternoon-night relationship we can give an estimate of the photometric night quality. In fact, in Section \ref{seeing} we have demonstrated a high correlation between the $FWHM^{Mean }_{Sat}$ and the $FWHM^{Mean }_{Ground}$ (see Tables \ref{fwhmsat}, \ref{fwhmsat1} and Figure \ref{cc}). In addition, in Section \ref{hda} we have shown how the afternoon data are correlated with the night data (see Tables \ref{forecast}, column 3).\\
With these two results we have a model that can provide a satellite seeing calculation and a forecast.
An interesting result are the values shown in Tables \ref{change} and \ref{change1}. The two tables show the monthly percentage of the changes in the observation conditions at the two sites during the 2009. We have obtained that at Paranal the variation of the meteorological conditions occur during the day time, but the months of February, March and April occur in the time interval between the 5 p.m. and 9 p.m. During the night the weather is almost stable. At La Palma we shown that the variation occur during the day.

\subsection{ACKNOWLEDGMENTS}

This activity is supported by the European Community (Framework Programme 7, Preparing for the construction of the European
Extremely Large Telescope, Grant Agreement Number 211257) and by Strategic University of Padova funding by title "QUANTUM FUTURE".\\
Most of data of this paper are based on the CLASS (Comprehensive Large Array-data Stewardship System).\\ CLASS is an electronic library of NOAA environmental data.\\ This web site provides capabilities for finding and obtaining those data, particularly NOAA's Geostationary Operational Environmental Satellite data.\\
Finally we acknowledge the Liverpool Telescope website staff.

\label{lastpage}

\end{document}